\documentclass[prb,twocolumn,superscriptaddress,showpacs,preprintnumbers,amsmath,amssymb]{revtex4}
\usepackage{amssymb}

\usepackage{graphicx}% Include figure files
\usepackage{dcolumn}%Align table columns on decimal point
\usepackage{bm}% bold math
\usepackage{epsfig}
\newcommand{\ignore}[1]{}

\begin{document}

\title{Graphene Switch Design: an Illustration of the Klein Paradox}

\author{Q. W. Shi}
\affiliation{Hefei National Laboratory for Physical Sciences at
Microscale, University of Science and Technology of China, Hefei,
Anhui 230026, People's Republic of China}

\author{Z. F. Wang}
\affiliation{Hefei National Laboratory for Physical Sciences at
Microscale, University of Science and Technology of China, Hefei,
Anhui 230026, People's Republic of China}

\author{Jie Chen}
\address{Electrical and Computer Engineering, University of Alberta, AB T6G 2V4, Canada}%

\author{Huaixiu Zheng}
\address{Electrical and Computer Engineering, University of Alberta, AB T6G 2V4, Canada}%

\author{Qunxiang Li}
\affiliation{Hefei National Laboratory for Physical Sciences at
Microscale, University of Science and Technology of China, Hefei,
Anhui 230026, People's Republic of China}

\author{Xiaoping Wang}
\affiliation{Hefei National Laboratory for Physical Sciences at
Microscale, University of Science and Technology of China, Hefei,
Anhui 230026, People's Republic of China}

\author{Jinlong Yang}\thanks{Corresponding author. E-mail: jlyang@ustc.edu.cn}
\affiliation{Hefei National Laboratory for Physical Sciences at
Microscale, University of Science and Technology of China, Hefei,
Anhui 230026, People's Republic of China}

\author{J. G. Hou}
\affiliation{Hefei National Laboratory for Physical Sciences at
Microscale, University of Science and Technology of China, Hefei,
Anhui 230026, People's Republic of China}

\begin{abstract}

An armchair graphene ribbon switch has been designed based on the
principle of the Klein paradox. The resulting switch displays an
excellent on-off ratio performance. Anomalous tunneling phenomena
are observed in our numerical simulations. According to our
analytical results, selective tunneling rule is proposed to explain
this interesting transport behavior. The switch design enriches the
phenomenon of the Klein paradox and also provides a platform to
study the paradox.
\end{abstract}


\maketitle

The recent fabrication of graphene has attracted a lot of research
interest.{\cite{1,2,3,4,5,6}} Graphene consists of a single atomic
layer of graphite, which can also be seen as a sheet of an unrolled
carbon nanotube. The edge carbon atoms of graphene ribbons have two
typical topological shapes: namely armchair and zigzag. Zigzag
graphene ribbon has a localized state near the Fermi level, which
originates from a gauge field produced by lattice
deformation.{\cite{7}} Such a localized state, however, does not
appear in armchair graphene ribbons. Armchair graphene ribbon can be
easily made to be either metallic or semiconducting by controlling
its width.{\cite{8}} This remarkable characteristic is very
attractive in making graphene-based nanoscale devices. Moreover, a
graphene ribbon is easier to manipulate than carbon nanotube (CNT)
due to its flat structure, and thus it can be tailored by using
conventional lithography techniques. The minimal conductivity
resulting from its infinite two-dimensional (2-D) graphene structure
and its gap-less energy band structure, however, limits the
performance of the 2-D graphene devices due to its poor on-off
ratio.\cite{9} Semiconducting graphene ribbon is expected to resolve
this on-off ratio problem.

Very recently, Geim's group proposed an experimental realization of
prediction of the so-called Klein paradox in a 2-D graphene
system.{\cite{9}} The Klein paradox predicts that the electron can
pass through a high potential barrier without exponential
decay.\cite{10} In this letter, we design a graphene ribbon switch
by utilizing such a paradox and study the associated tunneling
problem in a quasi-one-dimensional system. For this purpose, we
consider a semiconducting armchair junction connecting to a left and
a right metallic armchair graphene leads. Due to the band gap in our
junction, the electron within the energy gap is prohibited to
transmit through the junction without applied gate voltage, and the
graphene switch stays in the `off' state. The junction can be turned
`on' when an external gate voltage exceeds a threshold voltage. In
addition to the good on-off ratio performance of the proposed
switch, our design also provides a platform upon which to confirm
the Klein paradox. Unlike the 2-D graphene system, in which
electrons have to conserve their chirality in order to pass through
the barrier, in the quasi-1D graphene system, due to the confined
boundary, certain selective tunneling should be satisfied for the
resonant transmission of electrons.

\begin{figure}[htpb]
\begin{center}
\epsfig{figure=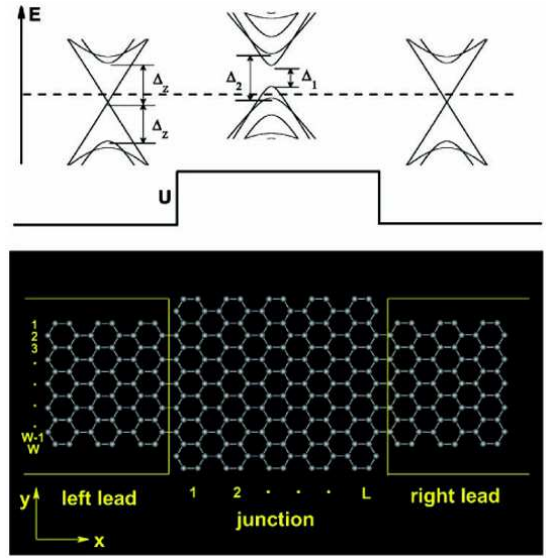,height=7.5cm}
\end{center}
\caption{(color online) Schematic diagrams of the energy band
dispersions and the proposed graphene switch.} \label{fig:fig-1}
\end{figure}

Our setup is shown in Fig.~\ref{fig:fig-1}, in which a
semiconducting armchair graphene is connected to two metallic
semi-infinite armchair graphene ribbon leads. The bottom subgraph of
Fig.~\ref{fig:fig-1} shows the atomic structure of the graphene
junction including three regions: left lead, middle graphene region,
and right lead. Both leads have the same width with $W=3m-1$ to
ensure that they are metallic. The width of the middle region is
chosen to be $W+4$ so that it is semiconducting.\cite{8} The top
subgraph in Fig.~\ref{fig:fig-1} shows the dispersion relation of
the left, middle and the right region of the switch, respectively.
The dashed line denotes the energy $E$ of incident electrons. $U$ is
the external potential applied to the middle graphene region. Unlike
conventional parabolic semiconductor energy band diagram (${\bf E}
\propto {\bf k}^2$), the energy of the lowest conduction subband in
the lead is linearly proportional to vector ${\bf k}$.{\cite{2}}
This dispersion relation indicates that carriers in this channel are
massless.

\begin{figure}[htpb]
\begin{center}
\epsfig{figure=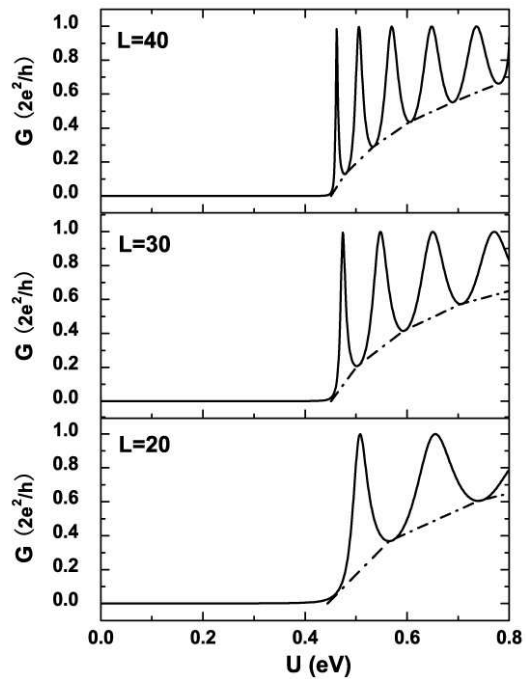,width=7cm}
\end{center}
\vspace{-0.2in} \caption{Conductance vs. the potential height in
the single-mode region with different length of the junction. Here
$W=23$ and $E=0.05eV$. The conductance oscillates above the envelope line
plotted with dash dot line.}
\label{fig:fig-2}
\end{figure}

The conductance $G$ of the switch is calculated by using the
Landauer-B\"{u}ttiker formula ($G$=$\frac{2e^2}{h}T$).{\cite{11}}
The transmission function $T$ is obtained by the recursive Green's
function method.{\cite{12,13}} In order to ensure the incident
electron energy lies within the single-mode region of the leads and
also in the gap of the middle graphene region, we choose
$E$$<$min($\frac{\Delta_1}{2},\Delta_{z}$). As shown in
Fig.~\ref{fig:fig-1}, $\Delta_{1}$ is the band gap between the
lowest conduction subband and the upmost valence subband of the
junction, $\Delta_z$ is the energy spacing between the bottom of
conduction subbands and the next subband within the lead. In the
following simulations, we set $W$=23 and thus obtain the
corresponding $\Delta_z$=0.65 eV, $\Delta_{1}$=0.38 eV and
$\Delta_{2}$=0.79 eV by the nearest neighbor tight-binding band
structure calculation.\cite{14} The calculated conductances are
shown in Fig.~\ref{fig:fig-2}. By applying an external gate voltage
at the middle graphene region, a potential barrier $U$ appears. In
general, if the chosen voltage potential makes the energy of
incident electron touch the first top of the valence subband of the
junction, the incident electron can easily pass through the junction
as expected and the switch should be turned `on'. However, our
simulating result shows that the incident electrons are almost
reflected completely and the conductance remains almost zero. This
phenomenon implies that the carrier has to satisfy an additional
condition to pass through the potential barrier.

When the potential barrier $U$ increases further, the switch turns
`on' and the first resonant transmission peak appears around
$U$=0.51 eV for $L$=20. This behavior can be easily understood by
the Klein's paradox. That is, with the help of the hole's channels,
the electrons can transfer through the large potential barrier
without exponential decay. The conductance oscillates above the
envelope line and increases as $U$ increases due to the resonance
and antiresonance transports, as shown in Fig.~\ref{fig:fig-2}. As
the length of the junction increases, the first resonant peak
becomes sharper and shifts towards left slightly and more resonant
peaks appear. These observations are easily understood by the
conventional resonance condition. However, to produce the first
conductance peak, the condition $U$$>$ $\Delta_{2}/2+E$=0.44 eV is
still required. This result strongly suggests that this peak results
from the second valence subband.

\begin{figure}[htpb]
\begin{center}
\epsfig{figure=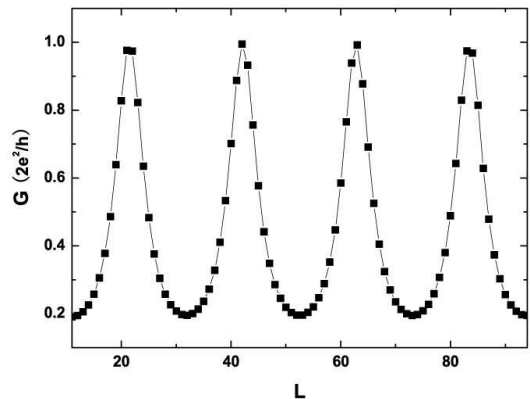,width=7.0cm}
\end{center}
\vspace{-0.2in} \caption{Conductance versus the length of the
junction with $W=23$, $E=0.05eV$ and $U=0.5eV$.}
\label{fig:fig-3}
\end{figure}

Fig.~\ref{fig:fig-3} shows the conductance as a function of the
junction length $L$ calculated by fixing $E$=0.05 eV and $U$=0.5 eV.
Such a choice provides two conducting channels in the middle region
for electrons passing through the junction. In general, the
conductance curve should be complicated with some glitches due to
the quantum interference between two channels.{\cite{13}}
Surprisingly, our numerical result exhibits a regular periodicity as
a function of $L$ in Fig.~\ref{fig:fig-3}. Based on the detailed
analysis of this conductance period with the conventional resonance
condition, we find that only the second valence subband provides a
channel to allow incident electrons pass, and the first valence
subband does not contribute. To verify this finding, similar
simulations were performed by changing the width of graphene ribbon
junction to $W=17$ and $W=20$ and the same phenomenon was observed.

Generally speaking, to pass through the middle region, electrons
have to satisfy conventional resonance condition which usually
describes the transfer of an electron through a junction via
resonant tunneling (T=1) due to the phase accumulation. It works
well in conventional semiconductor when the Schr\"{o}edinger
equation is used. In our switch model, the valence subbands in the
conductor can be shifted upwards with the applied external gate
voltage and provides conducting channels. When electrons satisfy the
convention resonance condition, $q_x$$L$=N$\pi$ with
N=0,$\pm$1,$\pm$2,..., here, q$_x$ denotes the x-direction (along
the propagation direction) component of wave-vector inside the
middle region, they can pass through the junction resonantly.
However, our simulation results show that the electron is still
prohibited to pass trough the middle region when the conventional
resonance condition is satisfied. To understand this anomalous
transport behavior more clearly, we derive an analytical expression
of the eigenfunction and eigenvalue of the perfect armchair graphene
ribbon with a finite width ($W$) as

\begin{eqnarray}
\psi_n^s(j,q_x)=\frac{sin(q_n\frac{\sqrt{3}}{2}aj)}{\sqrt{(W+1)\frac{\sqrt{3}}{2}a}}
\left(\begin{array}{c} 1\\ s\sqrt{\frac{\mu^*}{\mu}} \end{array} \right) \\
\varepsilon_n^s(q_x)=sV\sqrt{\mu\mu^*}
\end{eqnarray}

\noindent with
$\mu$=$2e^{iq_x\frac{a}{2}}cos(q_n\frac{\sqrt{3}}{2}a)+e^{-iq_xa}$,
$q_n=\frac{n\pi}{\frac{\sqrt{3}}{2}a(W+1)}$, respectively. Here, $a$
is the C-C bond length ($1.42${\AA}), $V$ is the nearest hopping
parameter (-3.0 eV), $n$ denotes different subband with 1, $\cdots$,
W, $j$ labels the atomic position in the y-direction with
1,$\cdots$,W, and $s$=1 and -1 describes the conduction subbands and
the valence subbands, respectively. The chirality of the electron in
the conduction subband or hole in the valence subband can be
determined by the good quantum number $q_x$ and $q_n$. To determine
the lowest conduction subband or the upmost valence subband, the
integer number n needs to satisfy the condition
$n$=$Nint[{\frac{2}{3}(W+1)}]$. Here the function of \emph{Nint}
rounds off the variable to a integer. To verify that electrons in
the lowest conduction subband cannot pass through the switch, we
calculate the corresponding transfer matrix element $P_{11}$, and we
obtain $P_{11}$=$|\langle s, n, q_x|\hat{V}|s', n^\prime,
q^{\prime}_x \rangle|^2$$\simeq$ $4.5\times10^{-14}$ eV$^2$ with
$s$=+1, $n$=16, $s^{\prime}$=-1, and $n^{\prime}$=19. Here,
$\hat{V}$=$V\sum\limits_{j} |j\rangle\langle j+2|$ with $j$=2,4,6,
$\cdots$,W-1 is the scattering operator coming from the sharp
interface between the lead and the middle region. In the elastic
scattering process in our system, the equation
$\varepsilon_n^{+1}(q_x)$=$\varepsilon_{n^\prime}^{-1}(q^{\prime}_x)+U$
has to be satisfied. Meanwhile, the applied voltage $U$ is large
enough to ensure the equation have a real numerical solution
$(q^{\prime}_x)$. The transfer matrix element $P_{12}$ between the
lowest conduction subband to the second upmost valence subband is
calculated to be around $0.27$ eV$^2$. In our switch the value of
these transfer matrix elements are almost independent of the width
of graphene ribbon based on our numerical simulations.

Unlike its 2-D counterpart,{\cite{10}} the sharp interface in our
design plays an important role. From our numerical results, it is
clear that the selective tunneling corresponding to two transfer
matrix elements ($P_{11}$ and $P_{12}$) can illustrate the
interesting transport process. When $U$$>$0.24 eV, the first valence
subband of the middle graphene junction is moved high enough to
provide a conducting channel, but $P_{11}$ almost equals to zero.
That is to say, no matter whether the chirality of electrons
conserves or not, the sharp interface in this symmetrical connection
prevents tunneling process. Electrons, therefore, are bounced back
and the conductance remains almost zero, as shown in
Fig.~\ref{fig:fig-2}. Increasing the length of the junction only
changes the conventional resonance condition, but the conductance
still remains zero. When we further increase the voltage potential
to $U$$>$0.44 eV, the second valence subband of the middle junction
moves upwards. The energy of incident electrons touches this channel
and the electrons can then tunnel through the potential barrier
without exponential decay.

\begin{figure}[htpb]
\begin{center}
\epsfig{figure=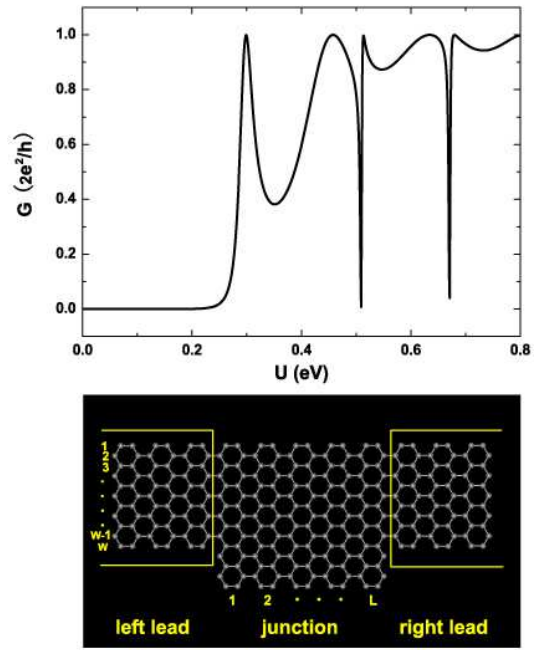,width=7cm}
\end{center}
\vspace{-0.2in} \caption{(color online) Top subgraph: conductance
vs. the potential height in the single-mode region. Bottom subgraph:
atomic structure of the T-shaped junction. Here $W=23$, $L=20$ and
$E=0.05eV$.}
\label{fig:fig-4}
\end{figure}

It should be pointed out that whether the first valence subband
contributes to the conductance or not depends strongly on the
geometric structure of the graphene ribbon junction. As an example,
a T-shaped junction with the same width of the previous symmetry
structure is shown in Fig.~\ref{fig:fig-4}. By choosing the same
parameters ($E$=0.05 eV, $W$=23 and $L$=20) as in the case of
Fig.~\ref{fig:fig-3}, numerical results in Fig.~\ref{fig:fig-4} show
clearly that the first conductance peak appears in the region where
the incident energy touches the first valence subband (0.24
eV$<$$U$$<$0.44 eV). The reason is that the transfer matrix element
($P_{11}$) has a finite value (about 0.24 eV$^2$) in T-shaped
junction. Our results suggest that the switch conducting behavior
can be manipulated by tailoring the graphene ribbon. In the range of
0.44 eV$<$$U$$<$0.7 eV, another interesting observation is that two
obvious conductance dips appear in the conductance curve. This
phenomenon can be attributed to the destructive interference effect
between two conducting channels.{\cite{15}}

\begin{figure}[htpb]
\begin{center}
\epsfig{figure=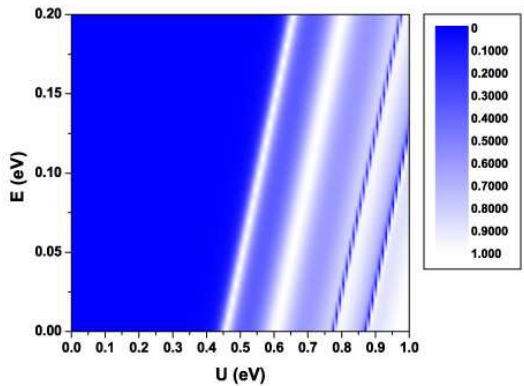,width=7.0cm}
\end{center}
\vspace{-0.2in} \caption{(color online) Three-dimensional plot of
conductance as the function of energy and potential height for the
configuration shown in Fig.1. Here $W=23$ and $L=20$.}
\label{fig:fig-5}
\end{figure}

To illustrate the switch effect more clearly in the single-mode
region, Fig.~\ref{fig:fig-5} presents a three-dimensional color
picture to display the conductance as a function of incident
electron energy (E) and the potential barrier height (U) for the
configuration shown in Fig.1. The color is scaled with the
corresponding conductance, white and blue colors correspond to $T$=1
(`on') and $T$=0 (`off'), respectively. It is clear that the figure
can be roughly divided into two regions bounded by the first peak
(the most left white line). At the left of the first peak (the first
white line), the switch stays shut off or all incident electrons are
bounced back. The junction begins to be turned `on' starting from
the right of the first peak by applying a certain threshold bias U
and its conductance oscillates with increasing $U$ as shown in
Fig.~\ref{fig:fig-2}. Although more channels can allow electrons
pass through or more resonance matching conditions can be satisfied
as $U$ increases, $T$ never exceed `1'. The reason is that the
energy of incident electron is limited within the single-mode region
in our calculations. Several additional interesting features can be
observed in Fig.~\ref{fig:fig-5}: (i) the peaks or the white lines
are parallel to each other and shift toward the right as the
incident energy increases. (ii) all peaks are straight lines
indicating the required voltage to open the switch increases
linearly as incident electron energy increases. (iii) the white
lines become wider and blue lines become narrower as $U$ increases.

In conclusion, the transport properties of a semiconducting graphene
ribbon sandwiched between two metallic graphene ribbon leads are
investigated. Switching behavior is observed according to our
numerical results. The junction has a good on-off ratio performance,
which is almost completely pinched-off without external gate voltage
and can be turned `on' by applying a threshold bias voltage. We find
that our numerical results are related closely to the Klein
phenomenon. Electron can pass through the junction when these Dirac
Fermions satisfy both the conventional resonance condition and the
selective tunneling rule. These findings are helpful for us to
construct and design graphene nanoelectronic devices in the near
future.

The authors would like to thank Michael Brett for his discussions
and support. We also would like to thank Shawn Sederberg and Woon
Tiong Ang for their assistance with the finalization of this paper.
This work is partially supported by the National Natural Science
Foundation of China, by National Key Basic Research Program under
Grant No. 2006CB0L1200, by the USTC-HP HPC project, and by the SCCAS
and Shanghai Supercomputer Center. Jie Chen would like to
acknowledge the funding support from the Discovery program of
Natural Sciences and Engineering Research Council of Canada (No.
245680).

\end{document}